\documentclass{desyproc}

\begin{document}
\title{Axion-electron coupling from the RGB tip of Globular Clusters}

\author{{\slshape  O. Straniero$^{1}$, I. Dominguez$^{2}$, M. Giannotti$^{3}$, A. Mirizzi$^{4}$}\\[1ex]
$^1$INAF-Osservatorio di Teramo, Teramo, Italy, and INFN sezione LNGS, Assergi, Italy\\
$^2$University of Granada, Granada, Spain\\
$^3$Barry University, Miami, USA\\
$^4$University of Bari, Bari, Italy, and INFN sezione di Bari, Italy}  

\contribID{Zioutas\_Konstantin}

\confID{16884}  
\desyproc{DESY-PROC-2017-02}
\acronym{Patras 2017} 
\doi  

\maketitle

\begin{abstract}
We present a preliminary study of the Globular Cluster RGB devoted to improve the 
available constraint for the axion-electron coupling. By means of multi-band IR photometry of
the cluster M3 we obtain $g_{ae}/10^{-13} < 2.57$ (95\% C.L.). 
\end{abstract}

\section{Introduction.}
Shortly after the central H exhaustion, the envelope of a Globular Cluster star rapidly expands 
up to a few hundred solar radii. Then the  star starts to climb the Red Giant Branch (RGB) and 
its luminosity progressively increases. Meanwhile, the He-rich core contracts, temperature 
and density increase, until electron degeneracy develops. Initially, the efficient conductive 
heat transport ensured by degenerate electrons makes the core almost isothermal. 
Later on, due to the central energy loss 
caused by plasma neutrinos, an off-center temperature maximum settle on. 
When the temperature rises above  the threshold for the He ignition, a thermonuclear runaway occurs
(He flash). 
This event coincides with the tip of the RGB.   
As firstly noted by~\cite{pakzy:1970mh} the luminosity of a RGB star 
essentially depends on the core mass. Then, as a consequence of the shell-H burning,
which is active at the bottom of the envelope, the core mass increases
and, in turn, the luminosity should increase.  
Therefore, the RGB tip luminosity
can be used to constrain the input physics that controls the growth of the He core 
mass during the RGB. In this framework a discrepancy between the observed RGB tip luminosity
and its theoretical prediction may be considered a hint of missed physical processes. 
In the following we will discuss the potential of RGB luminosity to constrain the coupling between
axions and electrons.
Like plasma-neutrinos, axions possibly produced in the core of a star which is 
climbing the red giant branch is an effective energy sink mechanism affecting the energy balance within the core 
and, in turn, the luminosity at the time of the off-center He ignition.  
The general rule is simple, the larger the production rate of weakly interactive particles 
produced by some thermal process the brighter the tip of the RGB. In this case the dominant 
axion production process is Bremsstrahlung, while Primakoff and Compton are suppressed because of the
high electron degeneracy.~\cite{viaux:2013a} make use 
of  I-band photometric data of M5, a well studied cluster of the Milky Way, to derive an upper bound 
for the strength of the axion-electron coupling: 
$g_{13}<4$ (95\% C.L.), or $g_{13}=2\pm2$. 
Here we present a project we started with the aim to improve this constraint. 
Instead of a monochromatic photometry of a 
single cluster, we will exploit multi-color near-IR photometric studies  of
a sample of galactic clusters from which bolometric magnitude can be directly derived and compared 
to the theoretical expectations. 
Brightest RGB stars are rather cool objects and their 
spectral energy distribution is dominated by near-IR light.  
In addition, to increase the statistical significance of the stellar sample our strategy is to 
combine high angular resolution
data of the crowded central regions (with HST or ground based adaptive optic telescopes),  
with large field photometries to cover the external regions of the cluster. 
The statistical significance of the stellar sample 
is an important issue 
for a correct determination of the RGB tip luminosity, a problem often 
ignored in previous studies. 
Our statistical approach to address this problem is discussed in the following section. 
To illustrate the potential of the method we will present the case of M3, another well studied
Cluster in the northern sky.     

\section{The observed brightest RGB star versus the RGB tip.}
\label{observed:tip}
The observed brightest star on the RGB does not necessarely coincide with the brightest 
point on the theoretical RGB evolutionary track or isochrone.
In principle, the probability  to observe the brightest RGB star 
as close as possible to the RGB tip depends on the 
total number of stars in the upper portion of the RGB. 
To estimate this probability, we make use of synthetic color-magnitude diagrams (CMDs, see~\cite{straniero:2015patras}).
In practice, we calculate a series of synthetic CMDs having the same input
parameters (age, metallicity and the like) and the same
number $N$ of RGB stars with bolometric magnitude in the range $m_{bol}^{tip}$ and  
$m_{bol}^{tip}+2$.  Although all the synthetic diagrams are computed with the same set
of input parameters, the $m_{bol}$ of the brightest star varies from CMD to CMD 
because of statistical fluctuations. In this way, for each $N$ we  calculate the probability 
density function (PDF) for $\delta m_{bol}$, which is  the difference between the $m_{bol}$ of the 
tip and that of the brigthest star.
Then, for each PDF (each $N$) we calculate the 
median and the standard deviation. 
As $N$ increases, the median approaches the mode, the most probable value, which is always $\delta m_{bol}=0$. 
In other words, the observed brightest star 
approaches the RGB tip as $N \rightarrow \infty$.   
Then, the absolute magnitude of the RGB is given by: 
$M_{bol}^{tip}=m_{bol}^{brightest~ star}-<\delta m_{bol}>-(m-M)_0 - A$,
 where $m_{bol}^{brightest~ star}$ is the apparent bolometric magnitude of the brightest RGB star,
$<\delta m_{bol}>$ is the median of the corresponding PDF, 
$(m-M)_0$ is the distance modulus and A is the extinction coefficient. 
Then, the total error budget is:
$\sigma_{obs}^2= \sigma_{stat}^{2} + \sigma_{d}^{2} + \sigma_{A}^{2} + \sigma_{ph}^{2} + \sigma_{BC}^{2}$
, where $\sigma_{stat}$ is the standard deviation of the appropriate PDF$_N$, and the other 4 uncertainties,
which represent the errors on distance, extinction, photometry and bolometric corrections,  
are obtained according to the available measuremts. 

\begin{table}
\label{clusters}
\centering
\begin{tabular}{|c|c|c|c|c|c|c|c|}
\hline
  cluster & [M/H] & $N$ & $<\delta m_{bol}>$  & M$_{obs}^{tip}$ &  $\sigma_{stat}$ & $\sigma_{obs}$ & $\sigma_{theory}$ \\
\hline
M3      & -1.16 &  125 &    0.045  &  -3.655  &   0.070 &   0.250 & 0.04 \\
\hline
\end{tabular}
\caption{Parameters used to estimate the magnitude of the RGB tip of M3. 
[M/H]$=log\frac{Z}{X}-log\left ( \frac{Z}{X} \right )_\odot$
 is the cluster metallicity and N is the number of stars within 2 mag from the tip. $\sigma_{theory}$
 includes the uncertainties due to the rates of the relevant nuclear processes}
 
\end{table}

As an example, in table \ref{clusters} we report the estimated value of the tip bolometric magnitude
for the cluster M3. The apparent bolometric magnitude of the brightest RGB star has been derived 
by ~\cite{valenti:2004MNRAS}, basing on a near-IR photometric dataset obtained by combining HST and 2MASS data.
In this case the major source of uncertainty is due to the distance.
 
\section{The theoretical RGB tip.}
\label{teory:tip}
Models of Globular Cluster stars have been computed by means of the FUNS code 
(for more details, see  ~\cite{straniero:2006NuPhA} and references therein). 
Our theoretical predictions for the RGB tip bolometric magnitude 
as a function of the cluster metallicity is well represented by the following relation:
  
\begin{equation}
\label{teotip}
M_{theory}^{tip}= 0.0161[M/H]^2 - 0.1716[M/H] - 3.87
\end{equation}

In the case of M3 we get $M_{theory}^{tip}=-3.65$, which is very close 
the observed one.
In general, uncertainties of the theoretical estimation of the RGB tip luminosity 
may be due to the main energy sources, such as the key nuclear reaction rates, or 
to the energy sinks, such as the plasma neutrino rates. 
The shell-H burning rate is controlled by the slower reaction of the CNO cycle, i.e., 
the $^{14}$N$(p,\gamma)^{15}$O reaction,
whose reaction rate has been directly measured by the LUNA collaboration down to 70 KeV~\cite{formicola:2004PhLB}. 
This limit is very close to the Gamow's peak energy for this reaction at the temperature 
of the shell-H burning of a RGB star. According to the STARLIB database,
we assume a $\pm$ 10\% uncertainty for this reaction. The corresponding  uncertainty for 
the theoretical tip bolometric magnitude is: $\sigma_{^{14}N(p,\gamma)^{15}O}=0.007$ mag. 
On the other hand, the start of the He burning, which coincides with the RGB tip, is controlled by the 
the $3\alpha$ reaction.
For T$\ge 100$ MK, the typical He ignition temperature, the uncertainty for the $3\alpha$ reaction 
rate is $\sim \pm$10\% ~\cite{fynbo:2005Natur}. This uncertainty implies 
an error for the estimated tip bolometric magnitude of $\sigma_{3\alpha}=0.0075$ mag. 
Note that the estimated error bars for the nuclear reaction rates do not include the uncertainty 
in the electron screening. 
The rate of plasma neutrinos production has been independently derived by several groups 
(\cite{haft:1994ApJ},~\cite{esposito:2002MPLA} and reference therein). 
This calculations commonly assume that the
neutrino dipole moment, $\mu$, is 0 (or negligible). A non-zero $\mu$ would enhance the neutrino
production rate, causing a more efficient energy sink and, in turn, leading to larger core-He masses and
brighter RGB tips~\cite{viaux:2013b}. Such an occurrence could explain or alleviate a 
discrepancy between stellar 
models and observed RGB tip luminosities, when the observed tip is brighter than the predicted one.
On the other hand, the same discrepancy may be solved by introducing an additional energy sink, such
as that induced by the production of non-standard weak interactive particles (e.g. axions).   
Keeping in mind this warning, in the following we will assume $\mu = 0$. Note that the
upper bound for $g_{ae}$ coupling constant we will obtain assuming $\mu = 0$ 
remains valid also in case of $\mu \neq 0$. This is not true for the hint 
we can get under the  $\mu = 0$ hypothesis.
Other model uncertainties are due to the adopted chemical composition, in particular,
the metallicity and the initial He mass fraction.
In the case of M3 we assume M/H$=-1.16 \pm 0.2$ and Y$=0.25 \pm 0.01$ that 
corresponds to  $\pm 0.035$ mag and $\pm 0.015$ mag on the RGB tip luminosity, respectively. 
Therefore, the total theoretical uncertainty is $\sigma_{theory}=0.04$.

\section{Axion-electron coupling from RGB tip}
\label{axion}
In this section we explore the hypothesis of an additional energy sink caused by the 
production of hypothetical bremstrahlung axions.
Therefore, we have computed models for different values of the axion-electron coupling constant,
namely $0 \leq g_{13} \leq 4$. The axion production rate has been computed according to
the prescriptions of~\cite{raff:1995PhRvD}, for low density plasma, and~\cite{nakagawa:1987ApJ},
at higher densities. Then, the inclusion of the axion cooling rate in the energy 
conservation equation leads to a larger core mass at the RGB tip and, in turn, to a larger luminosity. 
We obtain the following equation describing the relation among the RGB tip 
bolometric magnitude, the metallicity and the axion-electron coupling constant:

\begin{equation}
\label{teo2}
M_{teory}^{tip}= 0.0161[M/H]^2 - 0.1716[M/H] - 3.87 - 0.0239 g_{13}^2 - 0.078 g_{13}
\end{equation}
  
\noindent
which reduces to Eq. \ref{teotip} when $g_{13} = 0$.  
Then, the most probable value of $g_{13}$ is given by
the maximum of the likelihood function:
$L=A\exp\left [-(M_{theory}^{tip}-M_{obs}^{tip})^2/(\sigma_{theory}^{2}+\sigma_{obs}^{2})  \right ]$.
In the case of M3 we obtain $g_{13}=0.05$ with upper bound $g_{13}<2.57$ at 95\% C.L.. 
Because of the smaller difference between theory (no-axion) and observation, 
the upper bound we get for M3 is smaller than that obtained by~\cite{viaux:2013a} for M5.
A more substantial improvement of this bound may be 
obtained by combining data of more clusters, to increase the statistical significance of the sample, 
and increasing the accuracy of the distance determination, which is the major source of error.
 In this context the final data 
release of the astrometric satellite GAIA will produce a big impact ~\cite{pancino:2017MNRAS}.


\begin{footnotesize}

\end{footnotesize}


\end{document}